\documentclass[12pt]{article}
\usepackage{times,mathptm,moreverb,graphicx,epstopdf}
\begin{document}
\baselineskip=24pt
\bibliographystyle{unsrt}

\vspace{3in}

\begin{center}
{\Huge Entropic Trapping of Additive Particles at Polymer Surfaces and 
Interfaces}

\vspace{1in}

{ \large Galen T. Pickett }

\end{center}
\normalsize

\newcommand {\vol} {\bf}
\long\def\omit#1{}
\newcommand {\csi} {\xi}
\newcommand {\rhot} {\rho^{\prime}}
\newcommand {\csid} {\xi^{\prime}}
\newcommand {\gsim} {\stackrel{\textstyle >}{\sim}}
\newcommand {\lsim} {\stackrel{\textstyle <}{\sim}}
\newcommand {\vo} {v_o}
\newcommand {\vot} {v_o^{\prime}}
\newcommand {\aoverd} {a^{\prime}}
\newcommand {\peq} {d^3 P_{eq}}
\newcommand {\refto} {\cite}
\begin{center}
{\it   Department of Physics and Astronomy, 
California State University Long Beach,
	1250 Bellflower Blvd., Long Beach, CA 90840 \\

}
\end{center}

\vspace{0.5in}

\begin{center}
{\large\bf Abstract}
\end{center}
\baselineskip=24pt
I consider the possibility that 
Gaussian random walk statistics are 
sufficient to trap 
nanoscopic additives at either 
a polymer interface or surface.
When an additive particle goes to the free surface, two portions of the
polymer surface energy behave quite differently.
The purely enthalpic contribution increases the
overall free energy when the additive protrudes above the level of the
polymer matrix.
The entropic part of the surface energy arising
from constraints that segments near a surface can't cross it,
is partly relaxed when the additive moves to the free surface.
These two portions of the polymer surface energy determine the equilibrium
wetting angle formed between the additive and the polymer matrix, 
the measurement of which in an experiment would allow an independent
determination of each piece of the polymer surface energy.

\pagebreak
\hyphenation{mon-o-mer mon-o-mers homo-poly-mer homo-poly-mers co-poly-mer
	co-poly-mers an-i-so-trop-ic macro-molecules den-dri-mers
	Den-dri-mer}
\omit{
odell jeannie gibson, to get network running again
email 
campus hankee is working on it
}

\section{Introduction}
The addition of small, nanoscopic additives 
to polymer matrices is quite important
technologically and scientifically.\cite{bigfirst}
Even when the particles have no special properties other than hard
exclusions of polymer chains, they are useful in improving the
toughness of plastic components by preventing large-scale disruptions
of the matrix, for instance near a crack tip \refto{particle_tough}.
They can also pin local interfaces in a phase segregating polymer
blend, providing a mechanism for halting the growth of domains
and making blends behave more compatibly \refto{anna_blend_part}.

Using polymer patterns as templates for inorganic and metallic depositions
is another example  involving polymers and hard additive particles.
Here, microsegregation or other patterns in a polymer film can be selectively
decorated by the evaporation of metal \refto{pattern_deposition}.
I am particularly interested in these deposition processes
with
high surface-energy metal particles that reside at the free
surface of a polymer thin film \refto{hmj,shull_prl,find1,find2}.
Usually, patterned films cooled below their glass transition temperatures
are used for such purposes, with the goal of keeping the polymer
pattern static and allowing the additive particles to 
decorate the top of an impenetrable polymer layer \refto{glassy_decorate}.
The vicinity of the free surface of such patterned
films can often have a deep supression of $T_g$, giving the
metallic elements a chance to interact with a rubbery or molten polymer
layer \refto{bimetallic}.
However, even materials that would be expected to be wet with polymers in bulk
situations at temperatures well above the bulk value of $T_g$
seem upon deposition to remain pinned at the free polymer surface
rather than to penetrate to the interior of the films \refto{find1,pinned_free}.
When a further layer of polymer is floated upon these particles, these
trapped polymers are free to diffuse along the film normal.
When the floated layer is a different homopolymer, the particles can
remain trapped at the interface \refto{shull_trap_blend}.
When the particles wet neither of the polymer involved, they act as 
a compatibilizer, replacing polymer-polymer contacts with 
polymer-particle contacts \refto{anna_particle_compatibilize}.
\omit{While the particles in a deposition experiment will typically
decorate the material with the lowest particle-polymer surface
interaction, the fact that the 
particles 
are surface-trapped,
even when their bare surface energies are very large and the thermodynamic
penalty for failing to surround themselves with polymer segments
is prohibitive, they persist in remaining in the vicinity of the 
free surface.
}

To understand the fundamental physical effects driving these particles
toward free surfaces, 
the basic contributions to the bare polymer surface energy will
be crucial to understand.
These contributions have typically two parts, giving partial
surface energies
usually the same order of magnitude \refto{silberberg}.
Firstly, there is the enthalpic contribution to the polymer surface
energy, arising from short-range dispersion effects, contributing
$\gamma_p^o \equiv w_p /\sigma^2$ 
to the overall polymer surface energy.
Roughly speaking, every segment at the interface costs an interaction
energy $w_p$ compared to the reference of pure melt polymer.
Here, $\sigma$ is a typical polymer segment length scale.
Another interaction is $\gamma_{ps}^o \equiv w_{ps}/\sigma^2$ 
as well
as $\gamma_s^0 \equiv w_s/\sigma^2$, 
giving the enthalpic contribution to the surface energy between polymer and
the sphere, and the sphere and vacuum respectively.
Usually, $w_p \approx w_{ps}$, with both much smaller than the energy
scale for sphere-vacume contacts, $w_s$.
There is a strong thermodynamic driving force  keeping the
additive particle surrounded with at least a thin skin of polymer
segments \cite{find1,find2}.

The other main contribution to the polymer surface energy is entirely
entropic in nature, and arises from the constraint placed on a random walk
polymer configuration in the vicinity of an impenetrable interface
\refto{silberberg}.
Given a polymer segment near an interface, the adjoining polymer segments must
avoid crossing the forbidden surface.
This constraint can be thought of as an external bias applied to the chain
random walks, with the bias energy approximately equal to $k_B T$ per
constrained surface monomer.
Here, $k_B$ is Boltzmann's constant, and $T$ is the temperature.
Silberberg determined the correct weighting from an ingenious symmetry
argument, but the overall effect is entirely entropic, and contributes 
to the surface energy of $\gamma_e \equiv k_B T/\sigma^2$,
The total polymer surface energy, $\gamma_p$, 
is then given by $\gamma_p = \gamma_p^o + \gamma_e$.
Typically, $\gamma_p^o \approx \gamma_e$, so that entropic and enthalpic 
effects are equally important in determining $\gamma_p$.
It may be possible to separate these two effects, using one to
drive the creation of new surface area, while using the other
to stabilize the particle position.
Analyzing how these contributions affect the deposition or evaporation
of metallic particles onto a polymer melt is the main goal of this
paper.

As such,
the scenario of evaporating metallic particles onto a polymer film
is far from a new one, and has been studied in mean field models 
\refto{wetting_noolandi_ref_32} and interaction-site monte-carlo
simulations \refto{interaction_site}. 
Melt-particle interactions have been studied in
very detailed
bead-spring simulations \refto{bead-spring-nist}.
While the constraints placed on polymer conformations are dealt with
explicitly in an early mean field theory \refto{wetting_noolandi_ref_32},
that theory is most applicable to broad interfaces which dilutes the
importance of the segment-swapping contribution to the surface energy,
and therefore underestimates the entropic stabilizing effect.
Interestingly, this theory does predict that fully wet particles should
be stabilized near a polymer free surface at a range of approximately
the particle radius.
Studies aimed at studying the formation of nanoclusters of metallic 
particles after being embedded in the upper reaches of a polymer film
also show a marked tendency for the growing metallic particles to cluster
at the free surface as they grow, but it should be noted that this particular
simulation does not treat the polymer conformational entropy well, 
and therefore a layer of entropically active wetting segments is absent
in the equilibrated system.
The very through simulations in \refto{bead-spring-nist} investigate
the interactions of moderately long polymer bead-spring chains with
hard additive icosehedral particles. 
They find that polymer chains in contact with the
the surfaces of these particles
are indeed aligned along the particle surfaces,
indicating that Silberberg's notion of the
entropic contributions to the polymer surface energy are operating.

In a much simpler theory focusing on the Silberberg idea, I consider
impenetrable additive particles wetted
with polymer segments.\refto{find1,find2}
The question is {\em do such particles have an enhanced
surface activity}?
The quantitative prediction is that generally wetting
particles are surface active, and to a degree that will allow
the measurement of the relative size of the entropic and enthalpic
contributions to the system free energy.
\begin{figure}

\includegraphics[width=5in]{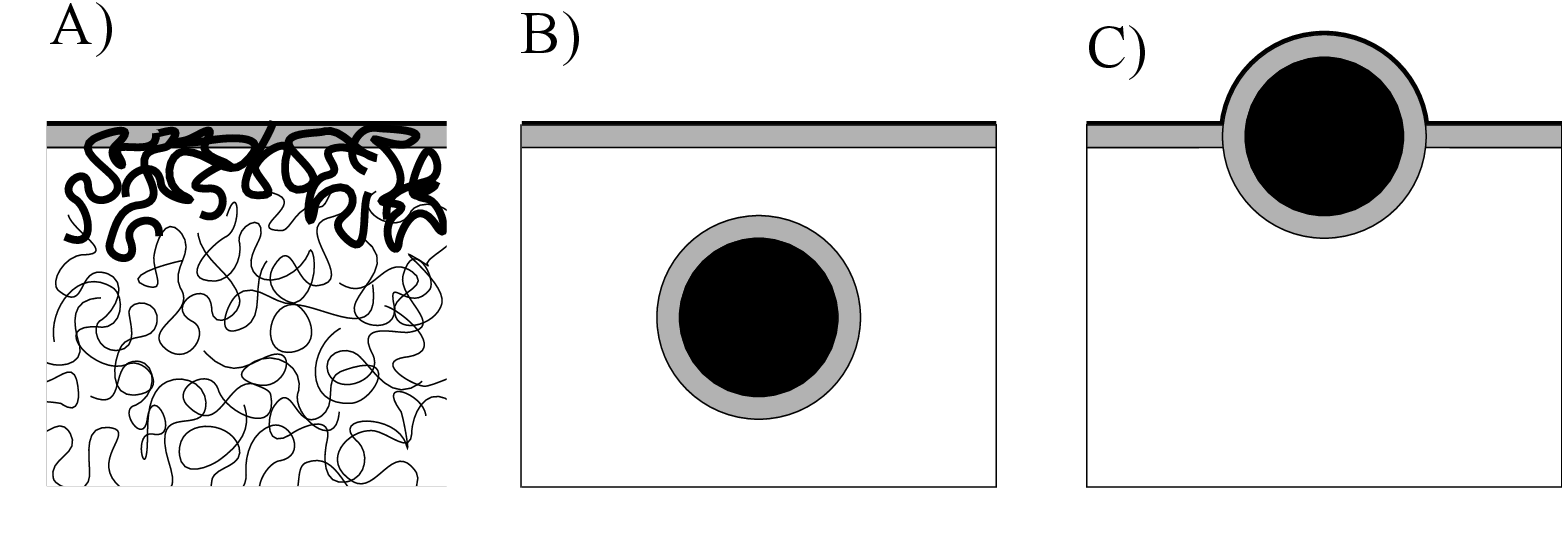}
\caption{ The surface of a typical
polymer melt. 
Chains with at least one monomer at the surface are shown
as bold lines.  
Each monomer within the shaded region at the free surface
represeents a free energy cost of $k_B T$ to maintain
the polymer conformations so that they do not cross the
free surface.
The bold line represents enthalpic, and the shaded 
region represents entropic, polymer-vacuum contacts.
A) bare melt, B) particle in the bulk, and C) particle surface stabilized by 
entropic contacts.
}
\label{fig.1}
\end{figure}
I will consider two simple, though related cases.
The first is a single spherical wetting particle in the 
vicinity of the free surface of a polymer melt, as in 
fig.~\ref{fig.1} part C.
The second case has a wetting, though partially
selective particle near the interface between incompatible polymer melts.
Even when the particles are preferentially wet with a single one
of the polymers involved, there are indeed situations in which
entropic gains keep the particles trapped
at the interface.
In the special case in which the spherical particles are deposited
upon a polymer melt, and then covered with an amorphous layer of 
the same polymer, the entropic pinning of the single particles is released, driving the particles away from the original interface,
as is observed experimentally \refto{shull_prl}.
It should be noted that the argument here is entirely equilibrium in character,
and does not depend on the presence of any other additive particles.
Thus, the effect described here does not depend on bridging interactions
between adjacent particles \refto{tom_bridge}.

After describing the effect for a single surface, I will treat the case
of a polymer-polymer interface.
After a short discussion, I will draw my conclusions.

\section{Free Surface}
As in fig.~\ref{fig.1} part A a polymer melt surface is characterized by a strong 
constraint on the conformations of all the polymer segments laying
within a polymer segment size, $\sigma$, of the interface.
Roughly speaking, for each such segment at the interface, the random
walk configuration of its chain must be biased so that both the preceding 
and the subsequent monomers on the chain are prevented from crossing
the interface.
In fig.~\ref{fig.1} part A the darkly drawn polymers are those which have at least one
monomer segment at the surface.
All of the monomer segments within the shaded region represent monomers
located at the surface, and each of these incurs an entropic cost of 
$k_BT$ in order to bias the random walk of its parent chain to respect
the interface.
Each of the segments at the free surface also are in contact with vacuum on
one side, and an semi-infinite sea of monomer on the other.
If the total surface area of the free surface of the melt is $L^2$,
then the free energy associated with maintaining this interface is:
\begin{equation}
F_{surf} = \frac{L^2}{\pi \sigma^2} k_B T \left(1 + \frac{w_p}{k_B T}\right)
\equiv L^2 \gamma,
\end{equation}
where $w_p$ is the energy to have a single monomer segment in contact
with vacuum, and $\gamma$ is the surface energy of the polymer.
Clearly
\begin{equation}
\gamma = \gamma_{entropic} + \gamma_{enthalpic},
\end{equation}
with
\begin{equation}
\gamma_{entropic}  \equiv \gamma_e =\frac{k_B T}{\pi \sigma^2} \mbox{ and }
\gamma_{enthalpic}  \equiv  \gamma_p^o =\frac{w_p}{\pi \sigma^2}.
\end{equation}
The free energy involving the presence of the additive
particle, a sphere of radius $R$ is:
\begin{equation}
F_{sphere} = \frac{4 \pi R^2}{\pi \sigma^2} k_B T \left( 1 + 
\frac{w_{ps}}{k_B T}\right).
\label{five}
\end{equation}
Here, $w_{ps}$ is the energy cost per unit monomer to make a
polymer-sphere contact.
In principle, there is another energy in the problem, $w_s$,
the energy needed to make a vacuum-particle contact for each polymer-segment
sized patch of surface.
We assume, as is generally the case when the additive particles
are metallic \cite{find2}, that 
\begin{equation}
w_p \approx w_{ps} \ll w_s
\end{equation}
and the additive particle are always wet with polymer segments.

Polymer segments near the sphere also endure entropic constraints.
Up to terms of order $\sigma/R$ this entropic cost is exactly the same for the 
polymer free surface, and represents the term in $F_{sphere}$ that
scales strictly as $k_B T$, as in eq.~\ref{five}.
Altogether, the total polymer $+$ sphere free energy is:
\begin{equation}
F_{in} = F_{surf} + F_{sphere} = L^2 (\gamma_{e} +
\gamma_{p}^o ) + 4 \pi R^2 ( \gamma_{e} + \gamma_{ps}^o),
\end{equation}
when the sphere is free to explore the interior of the matrix..

\begin{figure}
\includegraphics[width=5in]{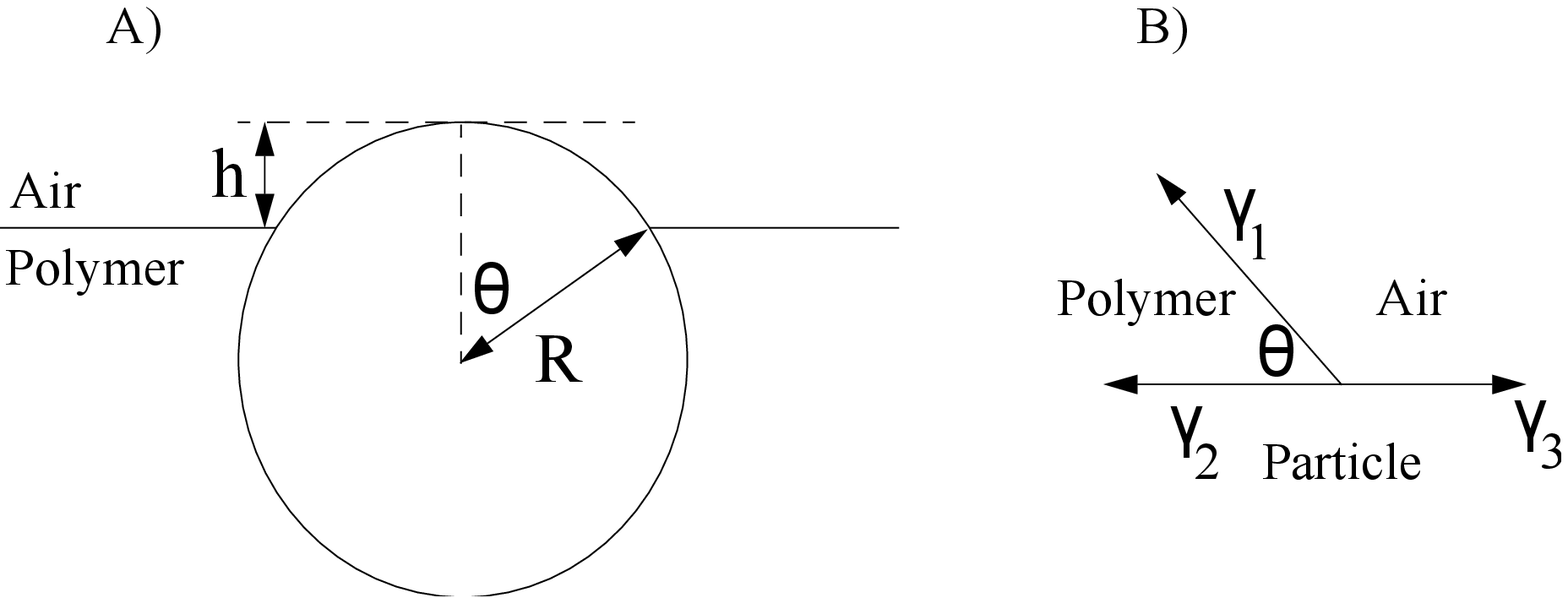}
\caption{ 
(A) Definitions of several quantities in the text are shown
here schematically.
This height of protrusion, $h$, the particle radius $R$ and
the wetting angle $\theta$ are shown.
(B) Young's law as applied at the polymer-metal contact.
}
\label{fig.2}
\end{figure}

Now, consider a situation as in fig.~\ref{fig.1} part C, where the sphere protrudes
some distance above the level of the polymer matrix.
For definiteness, let us suppose that the sphere protrudes a distance
$0< h < 2 R$ above the polymer matrix, so that the portion of the
sphere protruding covers an azimuthal angle $\theta$ as in fig.~\ref{fig.2}.
When $\theta =0$, the particle has just been brought to contact with the
free surface, and $h = R (1- \cos \theta)=0$.
When $\theta = \pi$, $h=2R$, and the particle has  been expelled
from the matrix.
The particle occupies a circular patch at the original interface
with a radius equal to $R \sin \theta$, so the contribution from the
surface of the unperturbed polymer matrix is:
\begin{equation}
F_{surf} = (L^2 - \pi R^2 \sin^2 \theta) (\gamma_e + \gamma_p^o),
\end{equation}
or, using elementary trigonometry,
\begin{equation}
F_{surf}(h) = (L^2 - \pi 2 h R + \pi h^2) (\gamma_e + \gamma_p^o).
\end{equation}
This represents the entropic and enthalpic contributions from the
flat polymer-vacuum interface.

The contribution to the system free energy  depending on the presence
of the sphere is also easy to estimate.
Using the fact that the sphere is still completely wet with polymer
segments, and that each of these segments still represents constraints
on the chain conformations of $k_B T$ per monomer, the 
entropic contribution to the sphere interaction is still {\em exactly
the same} as if it were completely immersed in the polymer melt.
The polymer-sphere contacts are exactly the same as well,
but there are now quite a few more polymer-vacuum contacts to 
account for.
In all:
\begin{equation}
F_{sphere}(h) = 4 \pi R^2 (\gamma_e + \gamma_{ps}^o) + 2 \pi R h \gamma_p^o
\end{equation}
Thus, the total free energy is 
\begin{equation}
F_{out}(h) = F_{surf}(h) + F_{sphere}(h),
\end{equation}
the free energy associated with having the particles a distance $h$ ``out''
of the polymer melt.

It only remains to determine the equilibrium value for $h$, which can be found
by minimizing $\Delta F(h) = F_{in} - F_{out}(h)$:
\begin{equation}
\Delta F(h) = (\gamma_e + \gamma_p^o) \pi  (h^2 - 2 h R) + 
2 \pi R h \gamma_p^o.
\end{equation}
This relation can be essentially read off from the schematic in 
fig.~\ref{fig.1} part C.
The dark line at the surface of the polymer melt marks polymer segment-vacuum
contacts, and the shaded region represents the volume in which entropic
biases are required to keep the chains from crossing
a surface.
There is less shaded area when the particle protrudes above the melt,
but with a cost in more polymer-vacuum contacts.

Minimizing $\Delta F$ with respect to $h$ yields the equilibrium value
for $h$:
\begin{equation}
h_{eq} = R \frac{ \gamma_e}{ \gamma_e + \gamma_p^o}.
\end{equation}
The wetting angle that the particle makes with the polymer matrix is
simply $\theta$, with an equilibrium value of:
\begin{equation}
\cos \theta_{eq}= \frac{\gamma_p^o}{\gamma_e + \gamma_p^o},
\label{youngs}
\end{equation} 
specifying the usual equilibrium contact angle.
Thus, the polymer coated particle acts just as Young's Law says it should,
as applied to the effective three-component interface consisting of the
polymer melt (with surface tension 
$\gamma_1=\gamma_p^o+\gamma_e$), the portion of the
particle immersed in the melt with an effective surface tension
$\gamma_2=\gamma_{ps}^o + \gamma_e$ and the portion of the particle protruding
from the surface of the melt with a surface tension of
$\gamma_3=\gamma_p^o +\gamma_e + \gamma_{ps}$.
According to Young's law, the wetting angle $\theta$ obeys:
\begin{equation}
\cos \theta = \frac{ \gamma_2 - \gamma_3}{\gamma_1},
\end{equation}
agreeing with the result in eq.~\ref{youngs}.

Thus, measurements of $h$ and $R$ and $\gamma$ disentangle
the two contributions to the {\em bare polymer surface energy},
and allows us to resolve its two parts.
In terms of the measurable quantities, we have 
\begin{eqnarray}
\gamma_p^o & = &  \gamma \frac{h}{R} \mbox{   (enthalpic), and}\\
\gamma_e & = &  \gamma \left(1- \frac{h}{R}\right) \mbox{   (entropic).}\label{observable}
\end{eqnarray}
Indeed, in the system of \refto{find1}, the protrusion of silver nanoparticles 
indicates that the $\gamma_e/\gamma_p^o$ is on the order of 2.
From the form of eq.~\ref{observable}, a number of interesting predictions
are possible.
First, the observable range of $h$ is $0 < h < R$, so that the additive
particle can either truly wet with the polymer, and have virtually no
surface activity ($h=0, \theta =0$) up to $h=R, \theta = \pi/2$.
The most deweting configuration possible here is when $\gamma_p =0$ so
that there is no enthalpic penalty for increasing $h$.
In this limit, we see that the additive particles act effectively as
{\em neutrally wetting} particles.
\omit{
These particles can not dewet with polymer, a conclusion depending upon
the assumption that bare particle-surface interaction is the largest surface
energy in the problem.  
When this breaks down, the additive can dewet in the usual manner.
}

For nonzero $h$, the amount of free energy pinning the particles
to the free surface can be estimated:
\begin{equation}
F_{pin} = \Delta F (h_{eq}) = - \pi R^2 \gamma 
\left|\frac{\gamma_e}{\gamma}\right|^2 = - \pi R^2 \gamma (1-\mu)^2.
\end{equation}
Clearly, when the enthalpic contribution to $\gamma$ dominates and 
$\gamma_e \ll \gamma$, (so that $\mu \approx 1$)
the energy pinning the spheres to the free surface can
become quite small, so that particle can readily leave the surface and
explore the interior of the melt.
If the melt is a film with thickness $H$, the translational entropy gained 
by the sphere when it leaves the surface and enters the films is
\begin{equation}
F_{trans} \approx \ln \frac{H}{R},
\end{equation}
so that the fraction of bound spheres vs. interior spheres can be
calculated for any given $H$ through:
\begin{eqnarray}
N_{surf} & =  & N_{tot} \frac{\lambda^2}{1 + \lambda^2} \\
N_{free} & \sim & N_{tot} \frac{1}{1+\lambda^2},
\end{eqnarray}
where 
\begin{equation}
\lambda = \frac{H}{R} \exp - \frac{F_{pin}}{k_BT},
\end{equation}
and $N_{tot}$ is the total number of deposited particles in the system.
In the discussion, below, I describe the basic sizes of these effects.
It should be kept in mind, however, that these particle
are driven toward the interface by an entirely entropic effect, which would
be removed if the the vacuum above the sample were replaced with
an identical layer of molten polymer.
Also, it should be kept in mind that I am interested here in the ultimate
equilibrium situation, and this argument completely ignores 
{\em how long} this equilibrium might take to be effected.
Long-lived metastable states can change the experimentally observable
phenomena (for example, the situation in which polymer
segments have not yet had enough time to envelop the additive
particle which is initially placed on the surface).
Also, this is a single-particle argument.
Interactions between the additives can lead to further surface-active effects
that have been discussed elsewhere \refto{shull}.

\section{Interface}
Now, I shall turn to the complimentary situation in which the additive
 particle
is placed at the interface between incompatible polymer melts,
$A$ and $B$.
The surface energy between these two polymers is $\gamma_{AB}$.
The bare particle surface energy is not relevant in this case, but we now
have two interactions between polymers and the particle:
$\gamma_{As}$ is the surface energy between polymer $A$ and the particle,
and likewise for $B$.
This energy has entropic and enthalipic parts as well:
\begin{equation}
\gamma_{As} = \frac{k_B T}{\sigma^2} + \frac{w_{As}}{\sigma^2}
\equiv \gamma_e + \gamma_{As}^o
\end{equation}
Similar quantities can be defined for $B$ polymer, and for the $AB$
interfacial tension.

There are two physically distinct situations to consider, the case in which
the particle wets completely with, say $B$ type monomers, and protrudes slightly into the upper $A$ region, and the more usual case in which the particle
is partially wet with both $A$ and $B$ monomers.
To determine which is the case requires a simple comparison of free energies.

When the particle partially wets with $A$ and $B$ monomers, I again define
$h$ to be the distance the particle protrudes into the $A$ region.
The free energy of the partially wetting parrticle is:
\begin{eqnarray}
F_{partial}(h) & = & (\gamma_{AB}^o + \gamma_e) \pi (h^2 - h R) + \nonumber \\
& &4 \pi R^2 \gamma_e + 2 \pi h R \gamma_{As}^o + \nonumber \\
& & (4 \pi R^2 - 2 \pi h R ) \gamma_{Bs}^o.
\end{eqnarray}
These terms correspond to the contribution from the $AB$ interface occupied
by the particle, the entropic surface energy of the particle, the
$A$-particle and $B$-particle surface contacts.
The equilibrium position of the particle, $h$, is found from minimizing
$F_{partial}(h)$:
\begin{equation}
h_{eq} = R \frac{ \gamma_{AB}^o - \gamma_A^o + \gamma_B^o}{\gamma_{AB}^o + \gamma_e} =
R \frac{\gamma_{AB} - \gamma_{As} + \gamma_{Bs}}{\gamma_{AB}},
\end{equation}
again a manifestation of Young's Law.
As usual, we can define a wetting parameter,
\begin{equation}
\epsilon = \frac{\gamma_A - \gamma_B}{\gamma_{AB}}
\end{equation}
so that $h_{eq} = (1- \epsilon) R$.
The free energy for this partially wetting case is thus:
\begin{equation}
F_{partial} = -\pi R^2 \gamma_{AB}(1-\epsilon)^2 + 4 \pi R^2 \gamma_{Bs}.
\end{equation}
Now,
assume that the particles are wet with
$B$-type fluid.
The free energy to maintain the particle protruding a distance
$h$ into the upper $A$ fluid is:
\begin{equation}
F_{B-wet}(h) = \gamma_{AB} \pi (h^2 - R h) + 4 \pi R^2 \gamma_{Bs} +
2 \pi h R \gamma_{As}^o.
\end{equation}
This represents exactly the same physical situation as is present
in Section 2, with the $A$ polymer playing the role of the vacuum.
Immediately, then, we have:
\begin{equation}
h_{eq} = R \frac{ \gamma_e}{\gamma_{AB}^o + \gamma_e}.
\end{equation}
The equilibrium free energy in this case is therefore:
\begin{equation}
F_{B-wet} = - \pi R^2 \gamma_{AB}(1-\mu)^2 + 4 \pi R^2 \gamma_{BS}
\end{equation}
where the appropriate wetting parameter here is
\begin{equation}
\mu = \frac{\gamma_{AB}^o}{\gamma_{AB}^o + \gamma_e}.
\end{equation}
Thus, there is a transition between the wetting and non-wetting situations
when 
\begin{equation}
(\epsilon-1)^2 = (\mu-1)^2,
\label{condition}
\end{equation}
the only relevant root of which is $\mu=\epsilon$.
That is, as long as $0<\epsilon<\mu$, the particle will be partially
wet, and $\mu< \epsilon<1$ will cause the particle to wet with $B$
monomers, although it is still confined to the $AB$ interface
by the entropic effect.
Clearly, when $\epsilon=0$, the equilibrium situation is that of neutrally
wetting particles, half covered with $A$ monomers, half covered with $B$
monomers, exactly straddling the $AB$ interface. 
As $\epsilon$ is increased, the particle moves toward the $B$ domain with
$h < R$, until a transition to the particle wet with $B$ monomers is achieved,
when the requirement of eq.~\ref{condition} is met.
\begin{figure}
\includegraphics[width=5in]{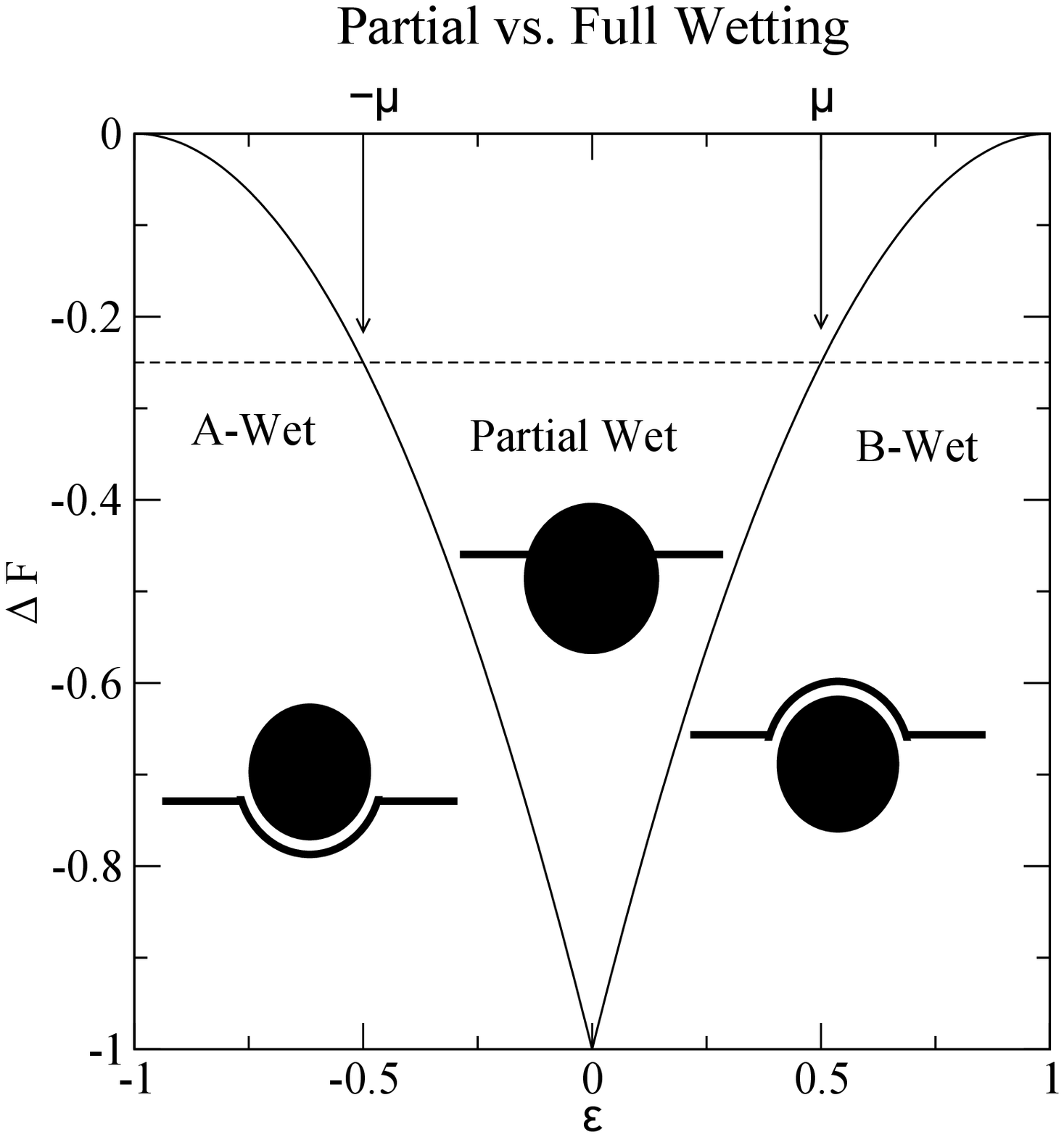}
\caption{ 
Given the entropic wetting parameter $\mu \equiv \cos \theta$
for the fully-$B$ wet sphere, if the partial
wetting parameter $\epsilon$  satisfies 
$|\epsilon| < |\mu|$, then the particle partially wets.  Otherwise, and
any further increase in $|\epsilon|$ is irrelevant for determining the energy binding the particle to the interface.
}
\label{fig.3}
\end{figure}
As in the case of particles residing at a free surface, an estimate can be
made of the equilibrium density of particles at the interface vs. the
equilibrium density of particles dispersed into the $B$ domain.
When $\epsilon<0$, the situation is exactly the same, except the transition
is toward the particle wetting with $A$ monomers, with $\gamma_{BS}
\rightarrow \gamma_{AS}$ in the reference energy.
The situation is shown schematically in fig.~\ref{fig.3}.

Thus, the sequence of events is as follows.
When high surface energy
particles are evaporated or deposited on the surface of a polymer
melt, they will first decorate themselves with a mass of restricted monomer,
creating a skin of $B$ monomer, say.
If the monomer-particle interaction is not strong (so that the entropic
energy dominates the equilibrium) the particles will be essentially
trapped at the surface, even upon annealing.
If more $B$ monomer is spun onto the system in a thick blanket, or if
a thin film is floated onto the existing $B$ surface, the entropic
stabilization of the particles toward the interface will disappear, 
and the particles will start to engage in restricted diffusion normal 
to the original interface (at possibly a very small rate, as the particles
in equilibrium resemble a star polymer, with many protruding 
loops and long arms of bound polymer).
If, on the other hand, a different type of monomer is deposited on the
surface, the particles may or may not still be trapped at the surface.
Very incompatible polymers ($\gamma_{AB}$ is very large) that neutrally
wet the particles ($\gamma_{As} \approx \gamma_{Bs}$) will be entropically
trapped at the interface.
The particles act as compatibilizers.
This state of affairs can continue up to and even past the point where the
particle wets with $B$ polymer.

\section{Discussion}
The scale of the energy trapping a typical particle at the free surface could
easily lead to practical trapping of even wetted particles.
The basic scale of the effect is $R^2 \gamma_{AB}$.
For a $R=10 nm$ gold sphere at a free polystyrene surface, this energy
scale is $ R^2 \gamma_{PS} \approx  (10 nm)^2 k_B T /(1 nm)^2$,
or approximately $100 k_B T$.
Therefore, a particle is stabilized by a factor of 
$100 k_B T (1-\epsilon)^2$.
For a typical flexible polymer, $\gamma_e \approx \gamma_p$, so that
$\epsilon \approx 1/2$.
Such a particle is therefore trapped by an energy of approximately
$25 k_B T$ per particle, and they protrude approximately $h=1/2 R= 5nm$
above the polystyrene surface.
Under normal circumstances, it would be virtually impossible to observe
such a particle leave the $A$ surface.
The particles are thus, for all intents, irreversibly attached to the
free surface, although their lateral motion is unrestricted.
As mentioned above, coating the sphere-polymer system with another
polystyrene layer will result in the particle being freed from the surface,
and engaging in diffusive motion along the film normal.

However, if a different polymer is spread on the surface of the
metal polymer system, we have to compare the wetting parameter,
$\epsilon$ for the $A-B$ particle system to the entropic wetting
parameter $\mu$ for the $AB$ interface.
Ideal conditions for unpinning of the spheres from the $AB$ interface
would require $\epsilon \approx \mu \approx 1$, 
and we can expect such conditions to occur regularly for very
selective interactions that are dominated by $w_{AS}$ and $w_{BS}$.
Note that the critical element here has nothing to do with
the incompatibility of the $A$ and $B$ fluids per-se, but is
sensitive to the interactions of the polymer with the
embedded sphere.

The entropic effect here stabilizing the particles at the interface
can be thought of an as extreme form of the depletion attraction of 
colloidal systems \refto{depletion}.
In that case, a solid sphere excludes the colloids from a spherical shell
of size $r$ around the sphere of radius $R$, where the colloid 
radius is $r$.
When two such spheres approach, the overlap of their exclusion shells
results in {\em fewer constraints} that need to be maintained in the
system.
In the current context, the colloidal particles are represented by the
polymer segments at the interfaces.
The particles sitting at the interface remove a circular patch of 
constraints that would otherwise have to be enforced, at the expense
of making energetically unfavorable contacts.
Thus, we can expect that spheres that have managed to escape the surface
and explore the interior of the film will be attracted by an effective
potential potential with range $\approx \sigma$ which counts up effectively
the number of redundant constraints when two spheres are in contact.
The size of the attraction is therefore
\begin{equation}
F_{attract} \approx \sqrt{ \frac{R}{\sigma}} k_B T.
\end{equation}
Thus, for the $10nm$ gold spheres considered above, the size of the attraction
is approximately $k_B T$, resulting in moderately sized clusters of 
particles, or indeed a much weaker adsorption of these particles at 
a solid-polymer interface.

The basic assumption here is that the polymer segments have sufficient time
to envelop a particle deposited atop a polymer film \cite{find1,find2}.
Even when the film itself is glassy, the segregation of free ends in the
melt \refto{kumar_surface} and the extra free volume expected at a free
surface can result in a significant reduction in both the glass transition
temperature and the effects of polymer chain topology \refto{glass_film}.

\section{Conclusion}
I have demonstrated an entropic attraction resulting from constrained
chain trajectories in the presence of a hard surface which can result
in the trapping of polymer particles at both a free surface and at
a polymer-polymer interface.
The size of the effect can easily be several tens of $k_B T$ per particle
for nanoscopic particles, and can even result in a few $k_B T$ attractive
short-ranged contact interaction between particles wholly within the polymer
matrix.

\pagebreak
\pagestyle{empty}
\omit{
{\bf \Large Figure Captions}

\noindent
\ref{schematic} {\bf Schematic.} (A) The surface of a typical
polymer melt. 
Chains with at least one monomer at the surface are shown
as bold lines.  
Each monomer within the shaded region at the free surface
represeents a free energy cost of $k_B T$ to maintain
the polymer conformations so that they do not cross the
free surface.
The bold line at the top of the melt represents 
enthalpic polymer-vacuum contacts.
(B) Here, a hard particle is inserted into the melt,
and is completely engulfed in the matrix.
The polymer chains themselves have not been rendered.
The shaded regions represent those monomers with entropic
costs associated with constraining the polymer segments
to respect the surface of the melt and the surface of the
particle.
Again, the dark line represetns polymer-vacuum contacts.
(C) The particle at the surface of the melt.
The particle is compeltely wet with a layer of monomers,
each of which represents an entropic cost of $k_B T$.
The solid line of polymer-vacuum contacts is now longer,
following the contour of the protruding particle.
Note that the overall shaded volume is lower when the
particle is resting at the interface, while the overal number of 
polymer-vacume contacts is larger. 
The balance of these two effects stabilizes the particle 
at the interface.

\noindent
\ref{schematic_2} {\bf Schematic.}
(A) Definitions of several quantities in the text are shown
here schematically.
This height of protrusion, $h$, the particle radius $R$ and
the wetting angle $\theta$ are shown.
(B) Young's law as applied at the polymer-metal contact.
$\gamma_1$ is the total polymer surface energy (including
both enthalpic and entropic contributions), 
$\gamma_2$ is the full metal-polymer surface interaction,
and $\gamma_3 = \gamma_{ps}^o + \gamma_e + \gamma_p^o$, 
representing the full polymer-metal surface interaction
along with the {\em bare} polymer-vacuume interaction.

\noindent
\ref{wetting} {\bf Wetting.}
Here, I show the determination between two wetting scenarios.
Given the entropic wetting parameter $\mu \equiv \cos \theta$
for the fully-$B$ wet sphere, if the partial
wetting parameter $\epsilon$  satisfies 
$|\epsilon| < |\mu|$, then the particle partially wets.
When $\epsilon$ is increased past $\mu$ the particle
wets with $B$ ($0<\mu$) or $A$ ($\mu <0$) monomers.
Any further increase in $|\epsilon|$ is irrelevant for determining the
wetting or the energy binding the particle to the interface.

\pagebreak
\begin{figure}
\includegraphics[angle=0,width=3in]{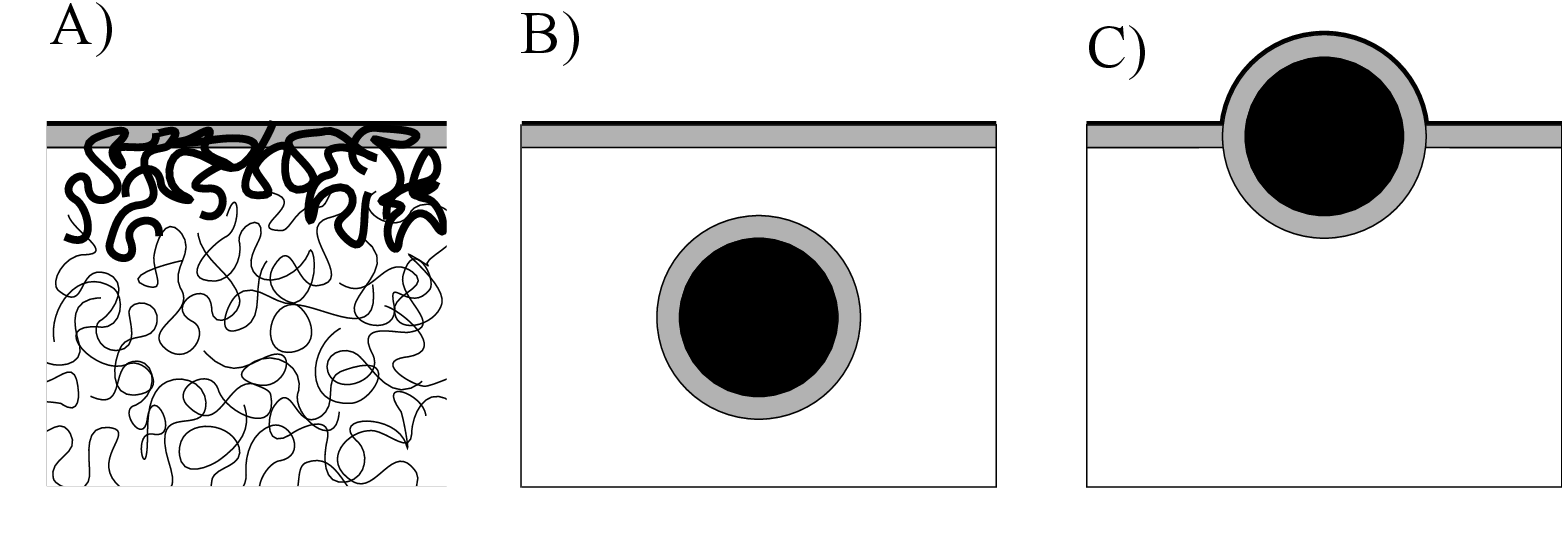}
\caption{Schematic.}
\label{schematic}
\end{figure}

\begin{figure}
\includegraphics[angle=0,width=3in]{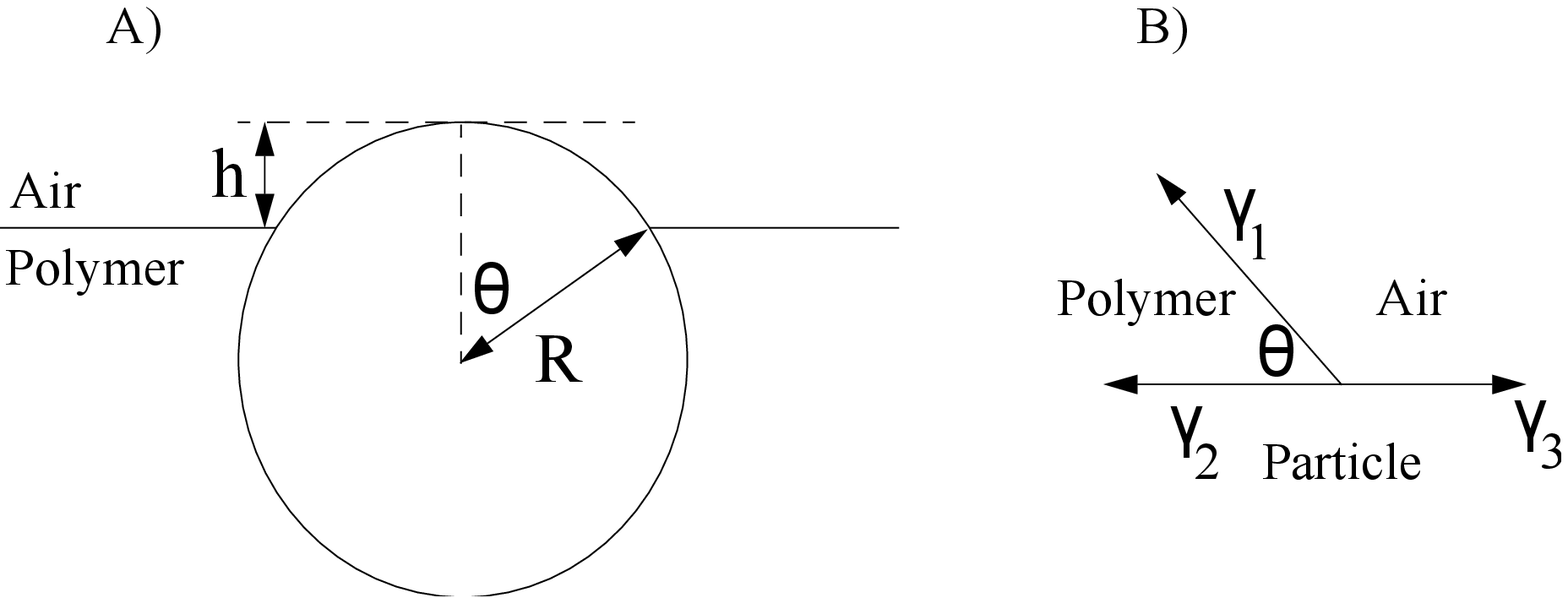}
\caption{Schematic.}
\label{schematic_2}
\end{figure}

\pagebreak
\begin{figure}
\includegraphics[angle=0,width=3in]{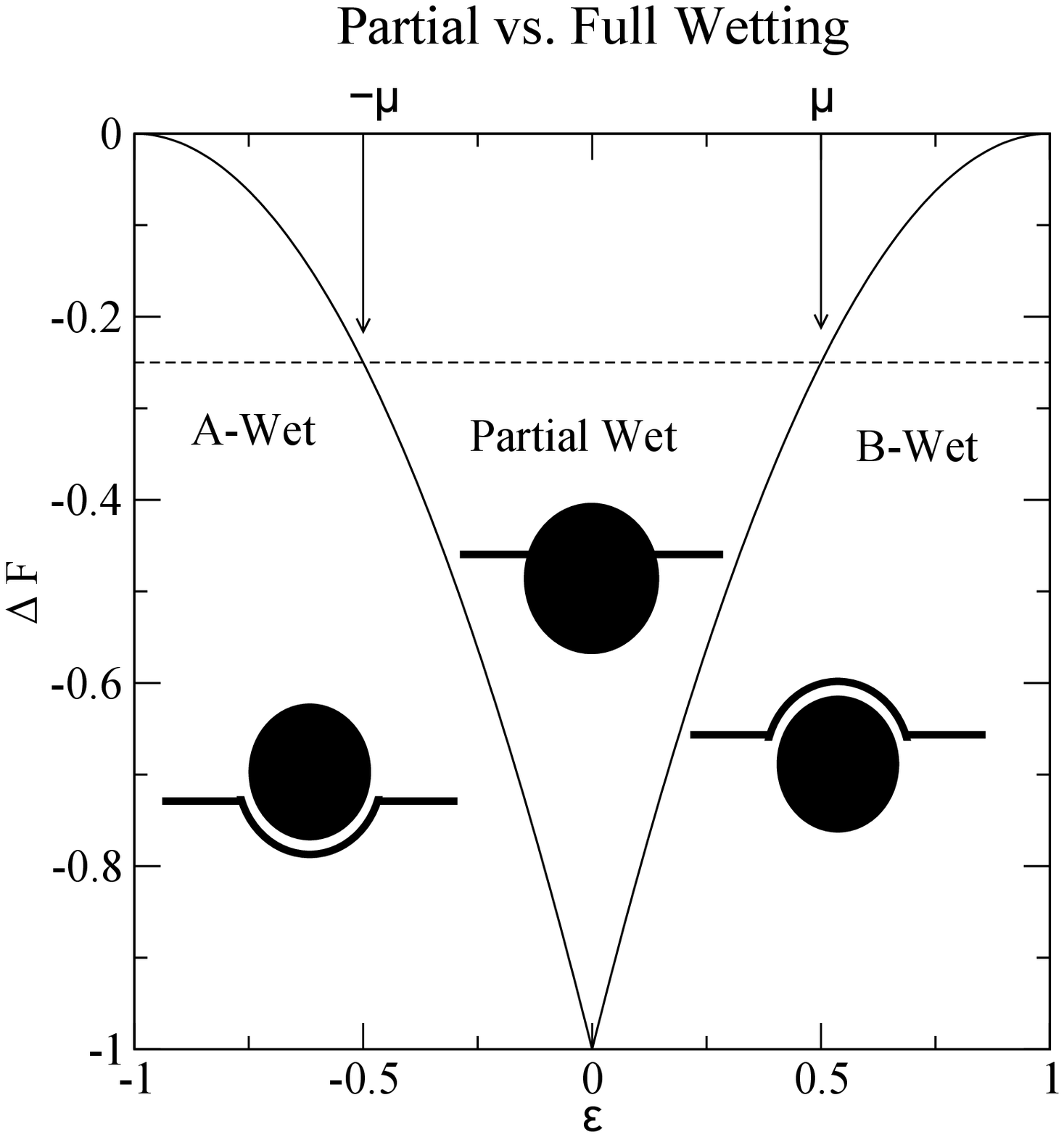}
\caption{Wetting}
\label{wetting}
\end{figure}
}
\omit{

}
\end{document}